\documentclass[twocolumn,superscriptaddress,showpacs,prl]{revtex4}

\usepackage{graphicx}
\usepackage{delarray}
\usepackage{amsmath}
\usepackage{bm}

\newcommand{\braket}[2]{\langle #1 \,|\, #2 \rangle}
\newcommand{\ket}[1]{| \, #1 \rangle}
\newcommand{\bra}[1]{ \langle #1 \,  |}

\begin{document}
\title{Quantum Cloning with Nonlocal Assistance:\\
Complement of Jozsa's Stronger No-Cloning Theorem
}

\author{Koji Azuma}
\email{azuma@appi.t.u-tokyo.ac.jp}
\affiliation{Department of Applied Physics, University of Tokyo, 
7-3-1, Hongo, Bunkyo-ku, Tokyo 113-8656, Japan}

\author{Masato Koashi}
\affiliation{Division of Materials Physics, Department of Materials Engineering Science, Graduate School of Engineering Science, Osaka University, Toyonaka, Osaka 560-8531, Japan}
\affiliation{CREST Photonic Quantum Information Project,
4-1-8 Honmachi, Kawaguchi, Saitama 331-0012, Japan}

\author{Hosho Katsura}
\affiliation{Department of Applied Physics, University of Tokyo, 
7-3-1, Hongo, Bunkyo-ku, Tokyo 113-8656, Japan}

\author{Nobuyuki Imoto}
\affiliation{Division of Materials Physics, Department of Materials Engineering Science, Graduate School of Engineering Science, Osaka University, Toyonaka, Osaka 560-8531, Japan}
\affiliation{CREST Photonic Quantum Information Project,
4-1-8 Honmachi, Kawaguchi, Saitama 331-0012, Japan}

\date{\today}

\begin{abstract}
We investigate the cases where a set $S$ 
of states $\{ \ket{\psi_i} \}$ cannot be cloned by itself, 
but is clonable with the help of another system prepared in 
state $\hat{\rho}_i$. When $S$ is pair-wise nonorthogonal, it is 
known that one can generate the copy from $\hat{\rho}_i$ alone, 
with no interaction with the original system. 
Here we show that a set containing orthogonal pairs exhibits a
property forming a striking contrast; For any such set, there 
is a choice of $\hat{\rho}_i$ that enables cloning only when 
the two systems are interacted in a purely quantum manner that 
is not achievable via classical communication.

\end{abstract}

\maketitle

It is impossible to deterministically make copies of nonorthogonal pure
states $\{\ket{\psi_i}\}_{i=1,\ldots,n}$, as stated in the no-cloning
theorem \cite{WZ82,Y86}.
This property suggests
that two quantum copies $\ket{\psi_i}\ket{\psi_i}$ 
are more ``informative''
than one copy $\ket{\psi_i}$, 
and it is natural to ask how much
more information the former has than the latter.
There have been quantitative approaches to this 
question, in which one considers quantities 
such as the 
optimal success probability 
in probabilistic cloning protocols \cite{DG98}, or
the optimal fidelity in 
approximate cloning protocols \cite{BH96,MPH97,GH97,BVOKH97,HB97,BDEMS98,GM97,BEM98}.
More recently, Jozsa proposed \cite{J02}
a qualitative approach
by asking
what kind of supplementary states
$\{\hat{\rho}_i\}_{i=1,\ldots,n}$ 
is required to make two copies $\ket{\psi_i}\ket{\psi_i}$
from the original state $\ket{\psi_i}$.
He found a striking property which he called 
the stronger no-cloning theorem:
For any pair-wise nonorthogonal (PNO) set of original states $\{\ket{\psi_i}\}_{i=1,\ldots,n}$,
whenever two copies $\ket{\psi_i}\ket{\psi_i}$ are generated with the help of
the supplementary state $\hat{\rho}_i$,
the state $\ket{\psi_i}$ can be generated from the supplementary state $\hat{\rho}_i$ alone,
independently of the original state,
namely,
\begin{equation}
\ket{\psi_i} \otimes \hat{\rho}_i \stackrel{{\footnotesize \mbox{CPTP}}}{\longrightarrow} \ket{\psi_i}\ket{\psi_i} \Longrightarrow  \hat{\rho}_i \stackrel{{\footnotesize \mbox{CPTP}}}{\longrightarrow } \ket{\psi_i},\label{eq:stronger}
\end{equation}
where CPTP stands for a completely positive trace-preserving map.
This result implies that
the original state 
is unable to provide even a partial help
in the creation of a copy, and hence 
the cloning process needs no interaction between the
original state $\ket{\psi_i}$ and the supplementary state $\hat{\rho}_i$.

While the above theorem only applies to PNO sets, 
the no-cloning theorem applies to a broader class.
We call that a set $\{\ket{\psi_i}\}$ is ``reducible''
iff we can divide the set into two nonempty sets
$S_1$ and $S_2$ 
such that any state in $S_1$ is orthogonal 
to any state in $S_2$.  
Since we can make a projective measurement to distinguish 
$S_1$ and $S_2$ without disturbing the original states,
we are allowed to consider only the irreducible sets 
in the problem of cloning. 
When the set of original states is irreducible but not
PNO, the cloning is still impossible but 
the stronger no-cloning theorem no longer applies.
Suppose that $\ket{\psi_1}$ and $\ket{\psi_2}$ are
 an orthogonal pair in such a set. As Jozsa pointed out \cite{J02}, 
we can take the supplementary information $\{\hat{\rho}_i\}$
such that $\hat{\rho}_1=\hat{\rho}_2$ and any other pair is orthogonal to each other.
In this case, the cloning is possible only if we combine 
the original state and the supplementary state. The 
required interaction between the two systems is purely
classical, namely, if the former system is held by 
Alice and the latter by Bob, classical communication 
between them is enough to accomplish the cloning.
This example might suggest a plausible interpretation
that the part of information held by an orthogonal 
pair is ``classical'', and it can help the creation 
of a copy by classically communicating with 
the system holding the supplementary information.

In this paper, we show that there are cases where
such an interpretation is not applicable, namely, there
are examples of original states $\{\ket{\psi_i}\}$
and supplementary states $\{\ket{\phi_i}\}$ 
that require quantum communication between Alice
and Bob to accomplish the cloning. The simplest
example is 
\begin{eqnarray}
    \ket{\psi_1}_A\ket{\phi_1}_B&=&\ket{0}_A\ket{0}_B,   \label{eq:ex1}\\
    \ket{\psi_2}_A\ket{\phi_2}_B&=&\ket{1}_A\ket{0}_B,    \\
    \ket{\psi_3}_A\ket{\phi_3}_B&=&2^{-1}(\ket{0}_A+\ket{1}_A)(\ket{0}_B+\ket{1}_B).\label{eq:ex3}
\end{eqnarray}
It is easy to see that if we apply a controlled-NOT gate between 
system $A$ (as control) and system $B$ (as target), 
we obtain the cloned state $\ket{\psi_i}_A\ket{\psi_i}_B$.
On the other hand, as we will prove later, Alice and Bob 
can never achieve the cloning through local operations
and classical communication (LOCC).
We further show that this example is not just a special 
case, but rather represents a general property shared by
{\em all} non-PNO irreducible sets. We prove that 
whenever the set of original states $\{\ket{\psi_i}\}_{i=1,\ldots,n}$ is 
irreducible but not PNO,
there always exists a set of supplementary states
$\{\ket{\phi_i}\}_{i=1,\ldots,n}$ such that the cloning process requires
quantum interaction between the two systems.

Throughout this paper, we assume that Alice holds
systems $A$ and $A'$, and Bob holds 
systems $B$ and $B'$.
System $A$ is secretly prepared in one of 
the original states $\{ \ket{\psi_i} \}_{i=1,\ldots,n}$,
and system $B$ is prepared in the corresponding 
supplementary state with the same index $i$ among the set 
$\{ \ket{\phi_i} \}_{i=1,\ldots,n}$. When Alice and 
Bob only communicate classically, difficulty of the 
cloning tasks depends on the requirement of who 
should possess the final copies. Since our aim here 
is to show the impossibility of the task, we adopt
the easiest task in which we place no restriction 
on the locations of the copies, as long as they 
are known after the protocol. More precisely,
we require that 
the task produces a classical outcome 
$XY$ which takes one of the three values $AA', AB, BB'$,
and the copies are produced accordingly as 
\begin{equation}
\ket{\psi_i}_A \ket{\phi_i}_B 
\stackrel{\mbox{{\footnotesize LOCC}}}{\longrightarrow}
\ket{\psi_i}_X
\ket{\psi_i}_Y\;(i=1,\ldots,n). \label{goal}
\end{equation}

We consider the following cases with three states:
\begin{eqnarray}
   \ket{\psi_1}&=&\ket{0},      \label{eq:ori1} \\
   \ket{\psi_2}&=&\ket{1},       \\
   \ket{\psi_3}&=&\alpha_0 \ket{0}+ \alpha_1 \ket{1}+\alpha_2 \ket{2}, \label{eq:ori3} 
\end{eqnarray}
where $\{\ket{i}\}$ is an orthonormal basis,
and $\alpha_0,\alpha_1,\alpha_2$ are real nonnegative numbers satisfying
$\alpha_0>0$, $\alpha_1>0$, and $\alpha_0^2+\alpha_1^2+\alpha_2^2=1$.
Note that this example essentially covers all non-PNO irreducible
sets of three states. 
For the supplementary states, we assume
\begin{eqnarray}
\ket{\phi_1}&=&\ket{0},  \label{eq:sup1} \\ 
\ket{\phi_2}&=&\left( \alpha_0 \alpha_1
+\sqrt{(1-\alpha_0^2)(1-\alpha_1^2) } \right) \ket{0} \nonumber \\
&&+\left( \sqrt{1-\alpha_0^2} \alpha_1-\alpha_0 \sqrt{1-\alpha_1^2} \right) \ket{1},   \\
\ket{\phi_3}&=&\alpha_0 \ket{0}+\sqrt{1-\alpha_0^2} \ket{1}.\label{eq:sup3}
\end{eqnarray}
Note that the case with $\alpha_0=\alpha_1=1/\sqrt{2}$
corresponds to the simple
example of Eqs.~(\ref{eq:ex1})-(\ref{eq:ex3}).

The states $\{\ket{\phi_i}\}_{i=1,2,3}$ have been chosen such that 
$\braket{\psi_i}{\psi_{j}}\braket{\phi_i}{\phi_{j}}=\braket{\psi_i}{\psi_j}^2$
for all $i$ and $j$. This relation assures that 
we can achieve the cloning task by a global operation, namely,
there is a unitary $\hat{U}$ such that 
$\hat{U}\ket{\psi_i}_A\ket{\phi_i}_B=\ket{\psi_i}_A\ket{\psi_i}_B\;
(i=1,2,3)$.
In fact, we can explicitly write down $\hat{U}$ as follows.
Let ${\cal H}_{\rm in}$ be the subspace spanned by 
$\{\ket{\psi_i}_A \ket{\phi_i}_B\}_{i=1,2,3}$, and  
${\cal H}_{\rm out}^{XY}$ be the one spanned
by $\{\ket{\psi_i}_X \ket{\psi_i}_Y\}_{i=1,2,3}$.
We construct an orthonormal basis of ${\cal H}_{\rm in}$
by Gram-Schmidt orthogonalization:
\begin{eqnarray}
\ket{v_1}_{AB}&:=&\ket{0}_A\ket{0}_B, \;\;
\ket{v_2}_{AB}:=\ket{1}_A\ket{\phi_2}_B, 
\nonumber \\
\ket{v_3}_{AB}&:=&(1-\alpha_0^4-\alpha_1^4)^{-1/2}
(\ket{\psi_3}_A \ket{\phi_3}_B
\nonumber \\
&&-\alpha_0^2\ket{v_1}_{AB}
-\alpha_1^2\ket{v_2}_{AB}),
\end{eqnarray}
and similarly for ${\cal H}_{\rm out}^{XY}$ as
\begin{eqnarray}
\ket{w_1}_{XY}&:=&\ket{0}_X\ket{0}_Y, \;\;
\ket{w_2}_{XY}:=\ket{1}_X\ket{1}_Y, 
\nonumber \\
\ket{w_3}_{XY}&:=&(1-\alpha_0^4-\alpha_1^4)^{-1/2}
(\ket{\psi_3}_X \ket{\psi_3}_Y
\nonumber \\
&&-\alpha_0^2\ket{w_1}_{XY}
-\alpha_1^2\ket{w_2}_{XY}).
\end{eqnarray}
Then, $\hat{U}$ is simply written as
$\sum_{i=1}^3 \ket{w_i}_{AB} {}_{AB}\bra{v_i}$.

Now we prove a lemma stating that Alice and Bob have to do 
a global quantum operation to achieve the 
cloning:

{\em Lemma 1} --- For the states of Eq.~(\ref{eq:ori1})-(\ref{eq:sup3}),
Alice and Bob can never achieve 
the cloning task of Eq.~(\ref{goal}) over LOCC.

The first step of the proof is to see what happens 
if Alice and Bob conduct the same cloning protocol 
with an initial state different from $\ket{\psi_i}_A\ket{\phi_i}_B$. 
Since $\braket{\psi_1}{\psi_{3}}\braket{\phi_1}{\phi_{3}}=\alpha_0^2>0$
and $\braket{\psi_2}{\psi_{3}}\braket{\phi_2}{\phi_{3}}=\alpha_1^2>0$,
the set $\{\ket{\psi_i} \ket{\phi_i}\}_{i=1,2,3}$ is irreducible.
This property allows us to determine
the output state for a general input state in ${\cal H}_{\rm in}$,
in the following way.
Suppose that after the cloning task of Eq.~(\ref{goal}),
we swap the states of systems $X$ and $A$ (if $X\neq A$),
and swap those of $Y$ and $B$ (if $Y\neq B$).
We further apply the unitary operation $\hat{U}^{-1}$
to systems $AB$. It is easy to see that the whole 
process does not alter the state of $AB$ when the 
initial state is one of the three states
 $\{\ket{\psi_i}_A\ket{\phi_i}_B\}_{i=1,2,3}$.
The property of such a disturbance-free process
is generally studied in \cite{koashi-imoto98,koashi-imoto00}.
In the present terminology, the result is stated as follows.
If the set of pure states $S:= \{\ket{\chi_i}\}_{i=1,\ldots,n}$ preserved 
in a process is irreducible, the process just leaves 
the subspace ${\cal H}$ spanned by $S$ as it is, namely, 
(a) any state $\ket{\chi}$ in ${\cal H}$ is preserved in 
the process, and (b) no information about the identity 
of the initial state $\ket{\chi}$ is revealed in the process. 
Applying this result in the present case, we see 
that
 (a') for any initial state $\ket{\chi}_{AB}$ in ${\cal H}_{\rm in}$,
the state of systems $XY$ after the cloning task 
is $\hat{U}\ket{\chi}_{AB}$ followed by swaps $X\leftrightarrow A$
 and $Y\leftrightarrow B$, and that
(b') the probability $\gamma_{XY}$ of the outcome $XY$
 is independent of the initial state. If we start from 
a general state $\sum_{i=1}^{3} \ket{v_i}_{AB}\ket{\mu_i}_R$
over $AB$ and an auxiliary system $R$, we should obtain 
the state $\sum_{i=1}^{3} \ket{w_i}_{XY}\ket{\mu_i}_R$
after the cloning task, and $XY$ is determined by 
a probability distribution $\gamma_{XY}$, which is 
fixed for the cloning task and independent of the 
initial state.

In what follows, we calculate the degree of entanglement 
of the output states when we try various input 
states over $ABR$. Using the fact that the entanglement 
never increases over LOCC, we derive bounds on 
the probabilities $\gamma_{XY}$, and show that 
the condition 
\begin{equation}
\gamma_{AA'}+\gamma_{AB}+\gamma_{BB'}=1 \label{eq:dete}
\end{equation}
can never be satisfied.

As a measure of entanglement, 
we use the entanglement monotone for 
a bipartite state $\ket{\Omega}_{AB}$  
defined by \cite{JP99}
\begin{equation}
E^{(l)}_{A;B}(\Omega):=1-\sum_{i=1}^{l-1} \lambda^{(i)},
\end{equation}
where $\lambda^{(1)},\lambda^{(2)},\ldots$ is 
the eigenvalues of $\mbox{Tr}_A[\ket{\Omega}_{AB}\bra{\Omega}]$
in the decreasing order,
and $l=2,3,\ldots$.
For each $l$, the value of $E^{(l)}_{A;B}$ never increases 
on average by LOCC. 
More precisely, 
a transformation of a state $\ket{\Omega^{\rm in}}_{AB}$ into 
$\ket{\Omega^{\rm out}_k}_{AB}$ with probability $\gamma_k$
can be done by LOCC iff 
\begin{equation}
E^{(l)}_{A;B}(\Omega^{\rm in}) \ge \sum_{k} \gamma_k E^{(l)}_{A;B}(\Omega^{\rm out}_k) \label{eq:LOCC}
\end{equation}
for any $l$ \cite{JP99}.

First we consider the case where system $R(=A'')$ is held by Alice
and try the following state as an input to the cloning process:
\begin{equation}
\ket{\Phi^{\rm in}}_{ABA''}=\frac{1}{\sqrt{3}} \sum_{i=1}^3 \ket{v_i}_{AB} \ket{i}_{A''},
\end{equation}
where $\{\ket{i}_{A''}\}$ are orthonormal.
Then, the process should produce the state  
\begin{equation}
\ket{\Phi^{\rm out}_{BB'}}_{BB'A''}=\frac{1}{\sqrt{3}} \sum_{i=1}^3 \ket{w_i}_{BB'} \ket{i}_{A''}
\end{equation}
with probability $\gamma_{BB'}$.
Since all $\{\ket{\phi_i}_{B}\}_{i=1,2,3}$ can be 
expanded by $\ket{0}_B$ and $\ket{1}_B$,
$E^{(3)}_{A''A;B}(\Phi^{\rm in})=0$.
On the other hand, $E^{(3)}_{A'';BB'}(\Phi^{\rm out}_{BB'})=1/3$. 
 From Eq.~(\ref{eq:LOCC}), we see that 
\begin{equation}
\gamma_{BB'}=0 \label{eq:gamma2}
\end{equation}
for any LOCC cloning process.

\begin{figure}[ttb]
  \begin{center}
    \includegraphics[keepaspectratio=true,height=37mm]{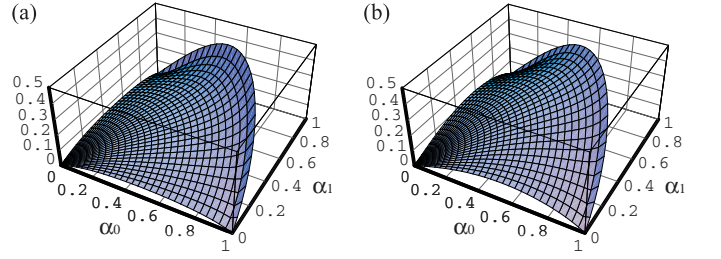}
  \end{center}
  \caption{(a) $E^{(2)}_{A;BB''}(\Psi^{\rm out}_{AB})-E^{(2)}_{A;BB''} (\Psi^{\rm in})$ and (b) $ E^{(2)}_{AA';B''}(\Psi^{\rm out}_{AA'})-E^{(2)}_{A;BB''} (\Psi^{\rm in})$ as a function of $(\alpha_0,\alpha_1)$.}
  \label{fig:ent1.eps}
\end{figure}

Next we consider another case with an input state
\begin{equation}
\ket{\Psi^{\rm in}}_{ABB''}=\frac{1}{2}\sum_{i=1}^2\ket{v_i}_{AB}\ket{0}_{B''}+\frac{1}{\sqrt{2}}\ket{v_3}_{AB}\ket{1}_{B''},
\end{equation}
where system $R(=B'')$ is held by Bob.
Then the output state of the cloning process is 
\begin{equation}
\ket{\Psi^{\rm out}_{XY}}_{XYB''}=\frac{1}{2}\sum_{i=1}^2\ket{w_i}_{XY}\ket{0}_{B''}+\frac{1}{\sqrt{2}}\ket{w_3}_{XY}\ket{1}_{B''}
\end{equation}
with probability $\gamma_{XY}$. 
As in Fig.~{\ref{fig:ent1.eps}},
numerical calculation shows that, for $\alpha_0>0$ and $\alpha_1>0$,
\begin{eqnarray}
E^{(2)}_{A;BB''}(\Psi^{\rm in})&<&E^{(2)}_{A;BB''}(\Psi^{\rm out}_{AB}), \label{eq:entI}
\\
E^{(2)}_{A;BB''}(\Psi^{\rm in})&<&E^{(2)}_{AA';B''}(\Psi^{\rm out}_{AA'}) \label{eq:entII}.
\end{eqnarray}
We can
also prove these inequalities analytically as follows.
Let us define marginal density operators for system $A$ as 
$\hat{\rho}^{\rm in}:=
\mbox{Tr}_{BB''}[\ket{\Psi^{\rm in}}_{ABB''}\bra{\Psi^{\rm in}}]$
and 
$\hat{\rho}^{\rm out}:=
\mbox{Tr}_{BB''}[\ket{\Psi^{\rm out}_{AB}}_{ABB''}\bra{\Psi^{\rm out}_{AB}}]$.
Eq.~(\ref{eq:entI}) is equivalent to the condition 
for the operator norms, $\|\hat\rho^{\rm in}\|>\|\hat\rho^{\rm out}\|$.
The difference between the two operators takes a very
simple form, $\hat{\rho}^{\rm in}-\hat{\rho}^{\rm out}
=\kappa(\ket{0}\bra{1}+\ket{1}\bra{0})$ with $\kappa>0$.
When $\alpha_2=0$, $\hat{\rho}^{\rm out}$ is 
also simply written in the form 
$p_0\ket{0}\bra{0}+p_1 \ket{1}\bra{1}$, and 
we see that $\|\hat\rho^{\rm in}\|>\|\hat\rho^{\rm out}\|$ holds.
When $\alpha_2>0$, we can easily confirm that all of the nine
matrix elements $\bra{i}\hat\rho^{\rm out}\ket{j}$ are 
strictly positive. 
Then, there is a state 
$\ket{\xi}:=\sum_j c_j \ket{j}$ with $c_j>0$
such that $\|\hat\rho^{\rm out}\|=\bra{\xi}\hat\rho^{\rm out}\ket{\xi}$
(Perron's theorem \cite{HJ85}). This leads to 
$\|\hat\rho^{\rm in}\|\ge 2\kappa c_0c_1+ \|\hat\rho^{\rm out}\|
>\|\hat\rho^{\rm out}\|$.

In order to prove Eq.~(\ref{eq:entII}), 
we consider 
$\hat{\rho}':=\hat{P}\hat\rho^{\rm in}\hat{P}$,
where 
$\hat{P}:=\ket{0}_A\bra{0}+\ket{1}_A\bra{1}$.
Since $\hat{\rho}'$ is represented by a $2\times 2$
matrix, 
it is 
tedious but not difficult to show that 
$\det[\hat{\rho}'-(1/2)\hat{P}]<0$ 
holds for $\alpha_0>0, \alpha_1>0$ \cite{proof}. 
This gives 
$E^{(2)}_{A;BB''}(\Psi^{\rm in})=1-\|\hat{\rho}^{\rm in}\|\le 
1-\|\hat{\rho}'\|<1/2$. Combining this with  
$E^{(2)}_{AA';B''}(\Psi^{\rm out}_{AA'})=1/2$,
we obtain Eq.~(\ref{eq:entII}).

Having proved Eqs.~(\ref{eq:entI}) and (\ref{eq:entII}),
the monotonicity Eq.~(\ref{eq:LOCC}) leads to 
\begin{eqnarray}
E^{(2)}_{A;BB''} (\Psi^{\rm in}) &\ge& \gamma_{AA'} E^{(2)}_{AA';B''}(\Psi^{\rm out}_{AA'})\nonumber\\
&&+ \gamma_{AB} E^{(2)}_{A;BB''}(\Psi^{\rm out}_{AB})\\
&>&(\gamma_{AA'}+\gamma_{AB}) E^{(2)}_{A;BB''}(\Psi^{\rm in}),
\end{eqnarray}
and hence
\begin{equation}
\gamma_{AA'}+\gamma_{AB}<1. \label{eq:gamma0,1}
\end{equation}
 From Eqs.~(\ref{eq:gamma2}) and (\ref{eq:gamma0,1}), 
Lemma 1 is proved.

A few remarks may be worth mentioning about Lemma 1. 
For simplicity, let us consider a deterministic 
transformation $\ket{\psi_i}_A\ket{\phi_i}_B\to
\ket{\psi_i}_A\ket{\psi_i}_B$.
Since $\braket{\phi_1}{\phi_2}>\braket{\psi_1}{\psi_2}(=0)$,
distinguishability in system $B$ alone 
between state $1$ and $2$ is improving in the process,
implying that the process may be regarded as 
 a kind of measurement on system $A$ that 
tries to distinguish between $\ket{0}_A$ and $\ket{1}_A$.
One might expect that in such a case, any 
superposition state between $\ket{0}_A$ and $\ket{1}_A$
will be destroyed in the process. 
That would be
surely true if the initial state of system $B$,
which we may regard as the
measurement apparatus, was
independent of $i$.
But the present case corresponds to
an atypical measurement in which the initial state of
system $B$ depends on $i$.
Then, 
rather surprisingly, there is a special initial 
state, $\ket{\phi_3}_B$, that enables the process
to keep a superposition state, $\ket{\psi_3}_A$,
unaltered. Lemma 1 means that this strange process,
extracting information while retaining a superposition,
can only be realized by interacting systems $A$ and 
$B$ in a purely quantum way.
 
Finally, we show that we can {\it always} find such
a subtle way of giving supplementary information
when the set of original states is irreducible but
not PNO.

{\em Theorem 1} --- 
For any 
non-PNO irreducible set $\{ \ket{\psi_i}_A
\}_{i=1,\ldots,n}$, 
there exists a set of supplementary states 
$\{\ket{\phi_i}_B\}_{i=1,\ldots,n}$ such that 
Alice and Bob can never achieve 
the cloning task of Eq.~(\ref{goal}) over LOCC.

{\em Proof.}
Let us call sequence 
$\ket{\Gamma_{1}},\ldots, \ket{\Gamma_{m}}$ a ``chain'' 
if $\braket{\Gamma_{i}}{\Gamma_{i+1}} \neq 0$ for $i=1,\ldots,m-1$.
First, we show that $S:=\{ \ket{\psi_i}_A \}$
includes a chain $\ket{\psi_{i_0}},\ket{\psi_{i_1}},
\ket{\psi_{i_2}}$ with $\braket{\psi_{i_0}}{\psi_{i_2}}=0$.
Let $\ket{\xi}$ and $\ket{\zeta}$ be a pair of orthogonal states 
in the non-PNO set $S$. Since 
$S$ is irreducible, it includes a chain $\ket{\xi}, \ket{\eta_1},
\ket{\eta_2}, \cdots, \ket{\eta_m}, \ket{\zeta}$ of length 
$m+2$ ($m\ge 1$). If $m=1$, this is the chain we seek. If 
$m\ge 2$ and $\braket{\xi}{\eta_2}=0$, we obtain the 
desired chain $\ket{\xi}, \ket{\eta_1}, \ket{\eta_2}$.
When $\braket{\xi}{\eta_2}\neq 0$,
 we can remove $\ket{\eta_1}$ from 
the chain and the remaining sequence of length $m+1$ 
still forms a chain connecting $\ket{\xi}$ and $\ket{\zeta}$.
Hence, repeating the procedure, we can always find a
chain
 $\ket{\psi_{i_0}},\ket{\psi_{i_1}},
\ket{\psi_{i_2}}$ with $\braket{\psi_{i_0}}{\psi_{i_2}}=0$.
Let us relabel the index $i$ in $S$ such that 
this chain becomes $\ket{\psi_{1}},\ket{\psi_{3}},
\ket{\psi_{2}}$. If we choose an appropriate basis,
these states are written as in Eqs.~(\ref{eq:ori1})-(\ref{eq:ori3}).
If we define the supplementary states 
simply
by $ \ket{\phi_i}$ of Eqs.~(\ref{eq:sup1})-(\ref{eq:sup3}) for $i=1,2,3$ 
and  $\ket{\phi_i}:= \ket{i-1}$ for $i=4,
\ldots, n$, the task of cloning 
becomes equivalent to the case with the three states 
considered in Lemma 1, and hence Theorem 1 is proved.

The present results, combined with the prior knowledge,
reveal the general property of quantum information in 
a set $S:=\{\ket{\psi_i}\}$ 
that manifests when one tries to clone it.
For the simple cloning, what matters is the reducibility 
of the set $S$. The reducible part is purely classical,
which is freely cloned. The irreducible part cannot be
cloned at all, which represents a quantum nature.
If one has an additional system with supplementary 
information $\{\hat{\rho}_i\}$, which has partial 
but not enough information to produce a copy 
$\ket{\psi_i}$ on its own, the class of 
irreducible sets are further divided into two types 
showing quite opposite behavior:
When the set $S$ is PNO, the original system is not helpful 
at all, and the cloning is still forbidden, which is 
the stronger no-cloning theorem \cite{J02}. When the set $S$ is 
not PNO (but irreducible), the original system can help
to achieve the cloning --- this fact itself is not surprising, 
since one may interpret that the orthogonal pairs of states in 
$S$ hold 
information in just a classical way. What is surprising is that 
we can always find an example of $\{\hat{\rho}_i\}$ such that 
this help is available only through a purely quantum operation
that is not achieved over LOCC. Hence the two cases, 
PNO and non-PNO, have properties which are both purely quantum
but are in a striking contrast with each other.

We have seen that the supplementary-state scenario
is very helpful in grasping the nature of quantum information
in a qualitative way. 
The scenario has also been combined with 
other protocols such as probabilisic cloning \cite{ASKI05}
and a novel cloning machine \cite{P99,Q05}.
We believe that we may also 
obtain a detailed quantitative understanding by
combining it with more elaborate protocols.

We thank J. Shimamura, T. Ohnishi, and N. Nagaosa for helpful discussions.
This work was supported by 21st Century COE Program by the Japan Society for the Promotion of Science and by a MEXT Grant-in-Aid for Young Scientists (B) No.~17740265.


\begin{thebibliography}{99}
\bibitem{proof}
We can explicitly rewrite as 
$(1-\alpha_0^4-\alpha_1^4)^2\times\det[\hat{\rho}'-(1/2)\hat{P}]=
-\sum_{k=0}^3 f_{2k}(x)(\alpha_0\alpha_1)^{2k}
-\sum_{k=0}^2 f_{2k+1}(x)(\alpha_0\alpha_1)^{2k+1}\sqrt{(1-\alpha_0^2)(1-\alpha_1)^2}$, where $x:=\alpha_0^2+\alpha_1^2$. $f_j(x)$ are all polynomials of $x$
and they turn out to be nonnegative for $0<x\le 1$. In particular, 
$f_6(x)=2>0$.
\end{thebibliography}
\end{document}